\documentclass{Interspeech2024}
\usepackage{booktabs}
\usepackage{multirow}
\usepackage{comment}
\usepackage{cite}

\interspeechcameraready

\title{Incorporating Class-based Language Model for Named Entity Recognition in Factorized Neural Transducer}

\name[affiliation={2,3}]{Peng}{Wang}
\name[affiliation={1}]{Yifan}{Yang}
\name[affiliation={1}]{Zheng}{Liang}
\name[affiliation={1}]{Tian}{Tan}
\name[affiliation={4}]{Shiliang}{Zhang}
\name[affiliation={1\dagger}]{Xie}{Chen}

\address{
 $^1$MoE Key Lab of Artificial Intelligence, AI Institute,  Shanghai Jiao Tong University, China \\
  $^2$Key Lab of Speech Acoustics and Content Understanding, Institute of Acoustics, CAS, China  $^3$University of Chinese Academy of Sciences, China 
  $^4$Alibaba Group}
\email{pengwang0104@gmail.com, chenxie95@sjtu.edu.cn}

\keywords{named entity recognition, factorized neural Transducer, class-based language model, beam search}

\newcommand\blfootnote[1]{
  \begingroup
  \renewcommand\thefootnote{}\footnote{#1}
  \addtocounter{footnote}{-1}
  \endgroup
}

\newcommand{\red}[1]{\textcolor{red}{#1}}

\begin{document}

\maketitle

\begin{abstract}
    
    

    Despite advancements of end-to-end (E2E) models in speech recognition, named entity recognition (NER) is still challenging but critical for semantic understanding. Previous studies mainly focus on various rule-based or attention-based contextual biasing algorithms. However, their performance might be sensitive to the biasing weight or degraded by excessive attention to the named entity list, along with a risk of false triggering. Inspired by the success of the class-based language model (LM) in NER in conventional hybrid systems and the effective decoupling of acoustic and linguistic information in the factorized neural Transducer (FNT), we propose C-FNT, a novel E2E model that incorporates class-based LMs into FNT. In C-FNT, the LM score of named entities can be associated with the name class instead of its surface form. The experimental results show that our proposed C-FNT significantly reduces error in named entities without hurting performance in general word recognition.
\end{abstract}
\blfootnote{$\dagger$ Corresponding author}

\section{Introduction}
\label{sec:intro}

End-to-end (E2E) models have become the de facto mainstream automatic speech recognition (ASR) due to their simplicity and promising performance \cite{li2022recent, he2019streaming, kim2019attention}. In ASR systems, the ability to recognize named entities, especially people names, is crucial for the semantic understanding of various downstream tasks. However, the joint optimization of acoustic and linguistic information in E2E models brings about difficulties in recognizing the long-tail named entities. This issue can be roughly attributed to two causes. One is the linguistic mismatch, as it is infeasible to cover all possible named entities in the training data. E2E models perform the speech recognition heavily relying on the training data and are inclined to assign a low probability to these long-tail or unseen name entities \cite{le2021contextualized,deng2021alleviating,winata2020adapt}. The other is the acoustic or pronunciation mismatch. Modern E2E systems normally adopt subwords derived from the spells, such as BPE and sentence piece \cite{sennrich2015bpe,kudo2018sentencepiece}. This is proven to work well for common words in English but found to perform poorly for named entities, especially for foreign names \cite{huang2020class}. In this paper, we mainly focus our discussion and research scope on the linguistic mismatch and leave the acoustic mismatch for future work.

There are consistent and active research efforts on improving named entity recognition in E2E models. Typically, an entity name list is prepared in advance and treated as contextual information. A popular practice is to apply various rule-based \cite{chen2019end,andresferrer21_interspeech,bruguier2022neural} or attention-based \cite{pundak2018deepcontext, sathyendra2022contextual,le2021deep} contextual biasing to facilitate the named entity recognition. Rule-based contextual biasing merely adds extra biasing weight when spotting some named entity in the partial recognition result. By contrast, the attention-based contextual biasing approach applies an attention mechanism over the named entity list, which implicitly boosts the probability of the named entity in the given list. According to the results reported in literature \cite{chen2019end,andresferrer21_interspeech,bruguier2022neural,pundak2018deepcontext, sathyendra2022contextual,le2021deep,chang2021context,jain2020contextual,williams2018contextual,zhao2019shallowfusion,munkhdalai2022fast,tong2023slot,fu2023robust}, the performance of named entity recognition can be improved significantly by applying contextual biasing. However, their performance might be sensitive to the biasing weight and degraded by excessive attention to the named entity list.  There are also some other attempts to enhance the named entity recognition in E2E models, such as the post-processing speller \cite{wang2022towards} and adding name tags \cite{huang2020class} into the target transcription during training.

Recalling the named entity recognition in conventional hybrid systems, where the acoustic model and language model are trained separately to capture the acoustic and linguistic information respectively, the class-based language models \cite{brown1992class,ward1996classlm} can be applied to compute the gross LM probability of a specific class, e.g. person name. This offers an elegant and theoretically sound solution to compute the LM probability of the named entity class instead of their surface form. However, in E2E models, the acoustic and linguistic information are fused to predict the next word jointly. As a result, there is no explicit language model component in standard E2E models \cite{graves2012sequence,battenberg2017exploring}, which hinders the direct use of class-based LM for named entity recognition. Recently, the factorized neural Transducer (FNT) \cite{chen2022factorized} was proposed to decouple the acoustic and linguistic information by introducing a standalone LM into the E2E models. In FNT, significant and consistent WER improvements can be achieved by improving the standalone LM based on text data.

Inspired by the success of class-based LM in hybrid systems and effective information decoupling in the factorized neural Transducer, we propose a novel E2E model, C-FNT, to integrate the factorized neural Transducer and class-based LMs. 
As a result, C-FNT performs named entity speech recognition in a similar way as hybrid systems with class-based LM while maintaining the key advantages of E2E models. The experimental results demonstrate that our proposed C-FNT presents significant error reduction in named entities without hurting performance in general word recognition.  It is also worth noting that, in this paper, we mainly focus on the person names as a case study for named entity recognition and choose the neural Transducer as the E2E model, given its popularity and excellent performance.


\section{Related Works}
\label{sec:format}
\subsection{Standard neural Transducer}
The standard neural Transducer \cite{graves2012sequence}, as shown in Figure \ref{fig:c-fnt}(a), consists of three modules, which are the acoustic encoder, label predictor, and joint network, respectively. The acoustic encoder consumes the acoustic feature sequence, and the label predictor takes the history subword sequence as input. The acoustic and linguistic representations are then combined in the joint network via a non-linear transform to predict the next output token, a vocabulary token or blank token $\phi$. In the neural Transducer, we aim to minimize the negative log probability over all possible alignments, which can be formed as follows:
\begin{equation}
\mathcal{J}_t=-\log P\left(\mathbf{Y} \mid \mathbf{x}\right)=-\log \sum_{\alpha \in \beta^{-1}(\mathbf{y})} P(\alpha \mid \mathbf{x})
\end{equation}
where the function $\beta$ is applied to convert the alignment $\alpha$ to the text sequence $\mathbf{Y}$ by removing the blank $\phi$.

The full integration of acoustic and linguistic information is beneficial for speech recognition, given the matched training and test scenarios. However, performance degradation is observed when there exists acoustic or linguistic mismatch during training and test time. As a result, E2E models are incompetent in named entity recognition as many named entities are rare words or even unseen words, especially foreign names in English conversation.

\subsection{Factorized neural Transducer}
\begin{figure*}[htb]
\centering
\includegraphics[width=1.0\linewidth]{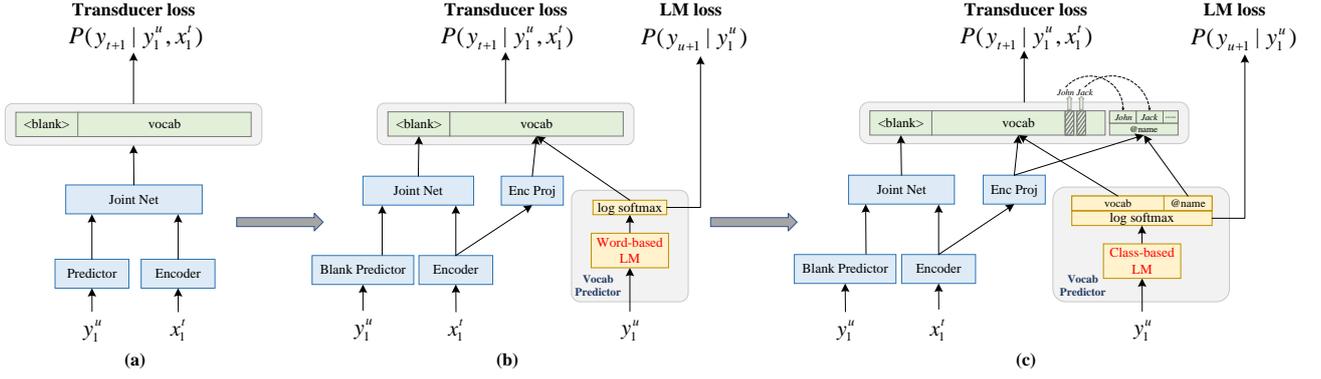} 
\caption{The illustration of three model structures: (a)standard neural Transducer (NT). (b) factorized neural Transducer (FNT). (c) proposed  factorized neural Transducer with class-based LM (C-FNT)}
\label{fig:c-fnt}
\vspace{-0.4cm}
\end{figure*}

The factorized neural Transducer (FNT) is an extension of the neural Transducer in which two predictors are adopted to predict blank and vocabulary tokens separately. The model structure can be seen in Figure \ref{fig:c-fnt}(b). The blank predictor in FNT is used to predict blank $\phi$, with a similar model structure and computation as the predictor in the neural Transducer. A standalone language model is introduced to serve as the vocabulary predictor in the FNT to predict the next vocabulary token. This factorization is found to be effective, and the improvement of the LM module by simply fine-tuning text-only data can be transferred to WER gains for speech recognition \cite{chen2022factorized}.  The effect of the LM module in FNT is similar to the role of LM in conventional hybrid systems.
The objective function of FNT is similar to the neural Transducer but with an additional LM loss. The loss function of FNT can be written as,
\begin{equation}
\mathcal{J}_f=\mathcal{J}_t-\lambda_f \log P_{LM}\left(\mathbf{Y}\right)
\end{equation}
Where $\log{P}_{LM}$ is the LM loss trained with cross-entropy, and $\lambda$ is a hyper-parameter to tune the effect of LM loss, which is set to 0.1 for simplicity in this paper. 

\subsection{Reviewing NER in neural Transducer and hybrid system} 
Before diving into the proposed approach, we explore the named entity recognition with the neural Transducer and explain why it performs poorly in NER. For clarity, an example sentence with person name \textit{``I will call Loretta Lynn''} is used for analysis. In most modern E2E systems, Byte Pair Encoding (BPE) \cite{sennrich2015bpe} is commonly used as the subword unit to reduce the vocabulary size and increase the generalization of rare words. This example sentence is then segmented to BPE sequence \textit{``I will call Lo \_retta Ly \_n \_n''}. In order to correctly recognize person names in the neural Transducer, the acoustic encoder and predictor need to jointly assign a high probability to this sequence given the corresponding speech. If there are not many sentences containing person name \textit{``Loretta Lynn''}, it is difficult for the model to produce this exotic BPE sequence \textit{``Lo \_retta Ly \_n \_n''}. Therefore, it poses a significant challenge for the E2E models to recognize such person names accurately. 


For the named entity recognition in a conventional hybrid system, the acoustic model maps acoustic features to the phone sequence, and the language model calculates the probability of any word sequence. Directly computing person names with word-level language models results in a similar long-tail issue, as discussed above. 
In order to resolve this issue, the class-based language model \cite{brown1992class} can be resorted to group person names into a specific class, denoting as \textit{@name}. Therefore, the LM probability,
\begin{equation}
\nonumber
    P_{LM} = P(Lo\ \_retta\ Ly \_n\ \_n|I\ will\ call)
\end{equation}
 can be computed with a class-based LM as
\begin{equation}
\nonumber
P_{CLM} = P(@name|I\ will\ call) * P(Lo\ \_retta\ Ly \_n\ \_n|@name)    
\end{equation}

where $P(Lo\ \_retta\ Ly \_n\ \_n|@name)$ is the prior probability of the specific name. A pretrained text-based NER model can be applied to convert person names to \textit{@name} on the LM training data. The class tag \textit{@name} can be modeled as common words under different contexts. Apparently, the class-based LM is more suitable for modeling the probability of named entities, and it receives widespread application in hybrid ASR systems.

\section{Proposed Approach}
\label{sec:format}
In this section, we present the proposed model, which is named C-FNT, aiming to incorporate the class-based LM into the factorized neural Transducer (FNT).

\subsection{Model architecture}
As shown in Figure \ref{fig:c-fnt}(b), there is a standalone word-based LM module in FNT. It is natural to replace this word-based LM with class-based LM, to empower this model with named entity recognition ability. The proposed C-FNT model is illustrated as Figure \ref{fig:c-fnt}(c). It can be seen that the major differences between FNT and C-FNT lie in the vocabulary predictor and the output layer. C-FNT adopts the class-based LM instead of the word-based LM in FNT, while for the output layer in C-FNT, an additional token is appended for the name class \textit{@name}. 

The construction of C-FNT is on top of FNT. 
The FNT model is first trained from scratch to factorize the acoustic and linguistic information. A class-based LSTM language model is trained separately on text data, which contains name tags with a NER model.
The class-based LM can then be plugged into the FNT by replacing the word-level LM module in FNT, which results in the C-FNT as shown in Figure \ref{fig:c-fnt}(c).

During the inference of C-FNT, the logits before softmax in the output layer consist of three types of nodes. The first two types are the blank $\phi$ and normal vocabulary BPE tokens out of the name class, which are computed the same as FNT; the last corresponds to the vocabulary BPE tokens within the name class, which need to be handled specifically.
Their acoustic and linguistic information are from different sources. The acoustic information is from the encoder projection as shown in Figure \ref{fig:c-fnt}(c), while the linguistic information is from the output of name class \textit{@name}  computed in the vocabulary predictor with the class-based LM.
The logit of the person name \textit{John} in the name class and the normal word \textit{John} out of the name class can be computed as below:
\begin{align}
    \textbf{z}_{@name}(\textit{John}|\textbf{y}_1^u, \textbf{x}_1^t) &= \log{P}_{CLM}(@name|\textbf{y}_1^u) + \textbf{z}_{enc}(\textit{John}|\textbf{x}_1^t) \label{eqn:logit_name} \\
    \textbf{z}_{voc}(\textit{John}|\textbf{y}_1^u, \textbf{x}_1^t) &= \log{P}_{CLM}(\textit{John}|\textbf{y}_1^u) + \textbf{z}_{enc}(\textit{John}|\textbf{x}_1^t) \label{eqn:logit_word}
\end{align}
It can be seen that they shared the same acoustic encoder output but with different vocabulary predictor outputs. 
Finally, the emitting probability at time $t$ can be computed with the softmax function, 
\begin{equation}
    P(y_{t+1}|\textbf{x}_1^t, \textbf{y}_1^u) = softmax([\textbf{z}(\phi), \{\textbf{z}_{voc}(w)\} ,  \{\textbf{z}_{@name}(w)\}])
\end{equation}
It is worth noting that in C-FNT, the probability is normalized over $2V+1$, instead of $V+1$, where $V$ is the vocabulary size excluding the blank $\phi$. At each time step, the score of the C-FNT is a vector with a size of $2V+1$   since each word might fall in the name class or out of the name class as normal words. However, the search within the name class is also constrained by the given name list, with only the names appearing in the name list can be emitted.

\begin{figure}[htb]
    \centering
    \vspace{-0.1cm}
    \centerline{\includegraphics[width=6.5cm]{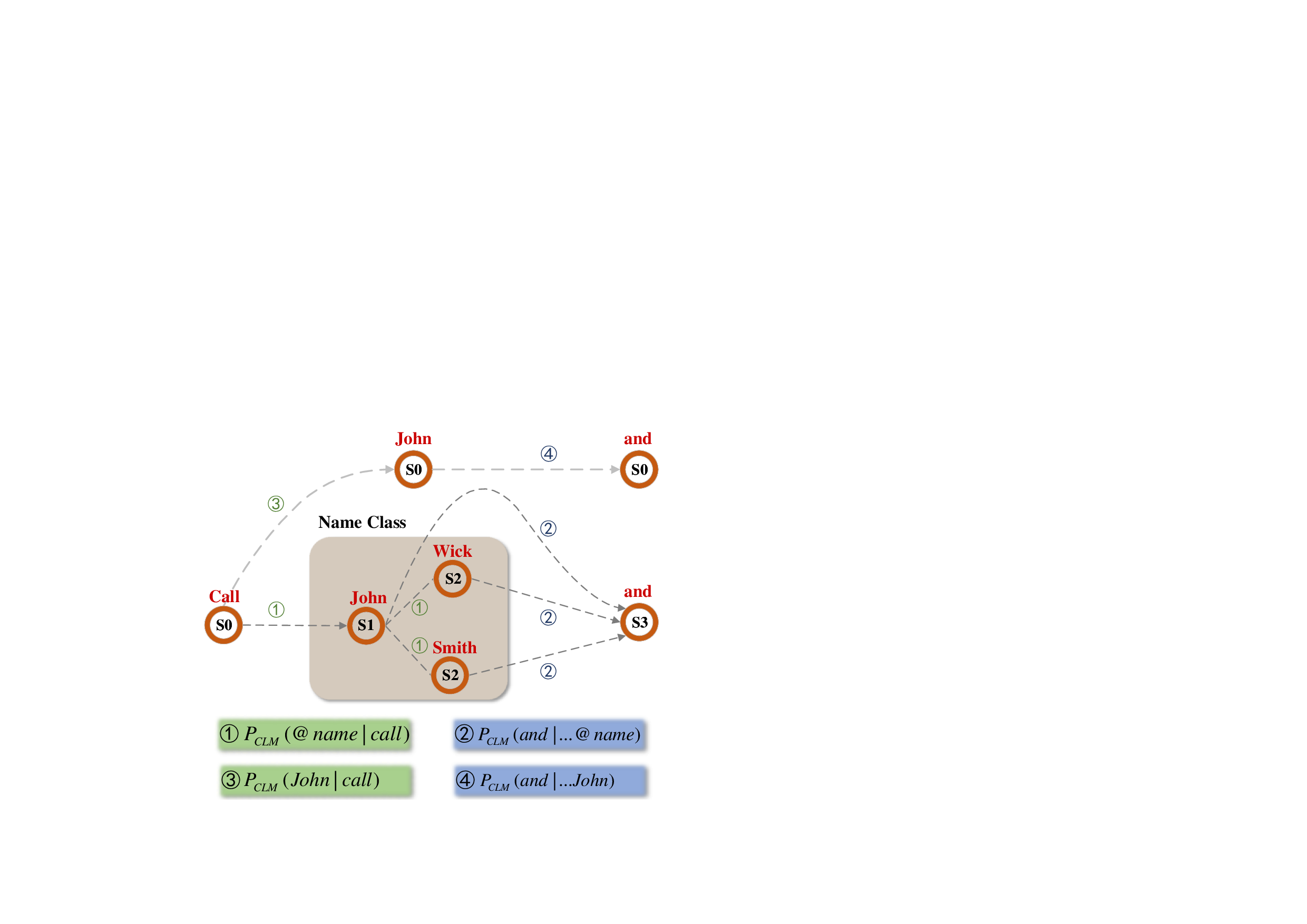}}
    \caption{An illustration of beam search decoding for C-FNT }
    \label{fig:c-fnt-beamsearch}
\end{figure}

\begin{table*}[!t]
    \centering
    \caption{Overall WER, Recall, Precision and F1-score results on Giga-name-test and NER-name-test test sets.}
    \label{tab:wer_ner_c_fnt}
   \begin{tabular}{c|c|c||c|c|c|c||c|c|c|c}
        \hline
        \multirow{3}{*}{Model} & \multirow{3}{*}{beam size} &\multirow{3}{*}{w/ name list}  & \multicolumn{4}{c||}{Giga-name-test} & \multicolumn{4}{c}{NER-name-test} \\
        \cline{4-11}
        &  &  & Overall &  \multicolumn{3}{c||}{Person name} & Overall &  \multicolumn{3}{c}{Person name} \\
        \cline{5-7} \cline{9-11}
          &  & & WER & Recall & Precision & F1-score & WER & Recall & Precision & F1-score  \\
        \hline
        FNT  &  5  & N & 15.8 & 54.4 & 56.9& 55.6 &  11.3   & 51.3 & 53.1 & 52.2 \\
        \hline
        \multirow{3}{*}{C-FNT} & 5 & N & 15.8 & 54.3 & 56.7 & 55.5 &  11.5   &  51.0  &  52.7  & 51.9\\
          &  5 & Y & 14.5 & 70.0 &72.1 & 71.0 &  10.7   &   66.8  &  68.6   & 67.7 \\
          & dynamic & Y & \textbf{14.4}   &  \textbf{71.5}   & \textbf{73.7}    &  \textbf{72.6} &  \textbf{10.6} & \textbf{69.5}&  \textbf{71.3} & \textbf{70.4}   \\
        \hline
    \end{tabular}
    \vspace{-0.4cm}
\end{table*}

\vspace{-0.4cm}
\subsection{Beam search decoding of C-FNT}
The decoding algorithms of the neural Transducer and FNT are the same since the model will output the same form of probability in spite of the differences in predictors. Greedy search and beam search \cite{graves2012sequence} can both be applied for decoding. However, for C-FNT, only beam search can be applied, while greedy search will cause severe performance degradation. The main reason is that during greedy search, only the best beam is kept. When this active beam enters the name class, only the person names specified in the name list can be generated. The limited feasible paths constrained by the entities in the name list might have a quite low probability. As a result, the model has to keep emitting blank $\phi$, and the decoding will be stuck in the name class. Therefore, for C-FNT, it is critical to adopt beam search and maintain active beams out of the name class, ensuring the speech recognition can be performed as word-level LM. 

During the beam search of C-FNT, each active beam can be classified into 4 different statuses based on its current and previous emitting vocabulary tokens.
\begin{enumerate}
    \item[S0] Normal status. Its current and previous tokens are both outside of the name class.  
    \item[S1] Entering status. Its current token is within the name class, and its previous token is outside of the name class.
    \item[S2] Staying status. Its current and previous tokens are within the name class.
    \item[S3] Exiting status. Its current token is outside of the name class, and its previous token is within the name class.
\end{enumerate}

If the statuses of all active beams are transiting between S0, this is the same as FNT decoding. When the status of one active beam jumps from S0 to S1, it enters the name class, and its logits are computed as Equation \ref{eqn:logit_name}. The name entity might be segmented into multiple BPE tokens. The transition from S1 to S2 means that the beam stays in the name class, and the LM status in the vocabulary predictor remains the same until it jumps out of the name class, corresponding to the transition from S2 to S3. The general process is illustrated in Figure \ref{fig:c-fnt-beamsearch}. It is also worth noting that, in order to avoid the decoding getting stuck in the name class and keep emitting blank $\phi$, we always keep at least one active beam search in the S0 status at each step. For the standard beam search decoding, the beam size is fixed and remains the same. In C-FNT, it may contain duplicated paths in all active beams as the paths are distinguished by the union of history word and status sequence. As shown in Figure \ref{fig:c-fnt-beamsearch}. The same LM history ``\textit{Call John and}'' corresponds to two different paths, as there are two different status sequences, which are ``S0$\rightarrow$S0$\rightarrow$S0" and ``S0$\rightarrow$S1$\rightarrow$S3". This duplication may squeeze other promising beams and cause performance degradation. Therefore we allow a dynamic beam size for beams within the name class. The LM states of the blank predictor in C-FNT are the same as those in FNT, which are based on word-level history.

\vspace{-0.1cm}
\section{Experiments and Results}
\label{sec:pagestyle}

\subsection{Datasets and Evaluation Metrics}
 We use the standard Gigaspeech dev\&test sets\cite{chen2021gigaspeech} to evaluate the general ASR performance. Two named entity test sets are used to evaluate the ASR and NER performance, the same as that used in \cite{liang2023improving}. The first test set is extracted from the Gigaspeech dev\&test sets, denoted as Giga-name-test set \cite{liang2023improving}. It is applied to assess performance in the in-domain condition since the acoustic training data is also from Gigaspeech. The second test set, referred to as the NER-name-test \cite{liang2023improving}, comprises about 1,800 sentences, and about $5.0\%$ of words are person names. It can be used to reflect the performance in out-of-domain scenarios. The detailed statistics of these two test sets can be found in Table \ref{tab:stats_ner_sets}. In addition, to train class-based LSTM LM, an open-source NER toolkit~\footnote{https://huggingface.co/dslim/bert-base-NER-uncased} is applied to the acoustic transcription of Gigaspeech-M set, and the detected names are then converted into a special class tag, \textit{@name}. These tagged sentences are concatenated with the original transcription. The class-based LSTM LM is trained on this augmented transcription.

For our experiments, WER is used to evaluate the general ASR performance, and the precision, recall, and F1-score of person names are the metrics to assess NER performance.
\vspace{-0.1cm}
\begin{table}[h]
    \centering
    \caption{Statistics of two NER test sets.}
    \resizebox{0.47\textwidth}{!}{
    \begin{tabular}{c|c|c|c|c}
        \hline
        \multirow{2}{*}{Test set} &  \multirow{2}{*}{$\#$sentence} & \multirow{2}{*}{$\#$word} & $\#$uniq & name  \\
         & & & name& percent \\
        \hline
        Giga-name-test & 1888  & 45k & 1434  & 8.3\%   \\
        NER-name-test  & 1845  & 57k & 1650  & 5.0\%\\
        \hline
    \end{tabular}
    }
    \label{tab:stats_ner_sets}
\end{table}
\vspace{-0.4cm}

\subsection{Experimental Setup}
\noindent\textbf{Baseline - Pretrained RNN-T.} The standard neural Transducer (RNN-T) is trained on the Gigaspeech-M set, consisting of 1000 hours of speech audio \cite{chen2021gigaspeech}. 5000 BPE tokens \cite{kudo2018sentencepiece} is derived from the acoustic transcription as vocabulary. The encoder consists of 12 Conformer \cite{gulati2020conformer} layers with 8-head attention of 512-dim and 2048 feed-forward hidden nodes. Two CNN layers are applied to subsample the input FBank features from 10ms to 40ms, which is then fed into the Transformer encoder to generate acoustic representation. The predictor contains two LSTM layers with a hidden layer size of 512. The RNN-T is trained from scratch on the Gigaspeech 1000h. The beam search algorithm in \cite{graves2012sequence} is used to decode and the beam size is 5.

\noindent\textbf{Baseline - Pretrained FNT.}  In FNT, both blank predictor and vocabulary predictor contain two LSTM layers with a hidden layer size of 512. The encoder and subsampling module are same with RNN-T. The FNT is also trained from scratch on the Gigaspeech 1000h and the model parameter is about 100M.

\noindent\textbf{Proposed C-FNT.} About 16.5k sentences sampled from Gigaspeech-M set are tagged with names. C-FNT is constructed by fine-tuning the LSTM word-based LM of FNT on these tagged transcription. The proposed C-FNT has a similar number of model parameter with FNT at about 100M. During beam search decoding, the name list is given to the model and the beam search decoding algorithm described in section 3.2 is used. As to beam size, two different scenarios are considered. In order to make a direct comparison with FNT, beam size of 5 is used first. Additionally, we also conduct experiments with dynamic beam size, expecting to avoid beam duplication when decoding in the name class.

\noindent\textbf{Name List Configuration.} For the Giga-name-test and NER-name-test sets, the sizes of the name list are 1434 and 1650, respectively. In addition, we want to test whether the proposed C-FNT could maintain its original performance given no name list. Therefore, we first decode C-FNT with no name list. We refer to this configuration as \textbf{empty name list}.

\vspace{-0.1cm}
\begin{table}[h]
    \centering
    \caption{Overall WER results on Gigaspeech dev\&test sets given \textbf{empty name list}.}
    \label{empty name list test}
    \begin{tabular}{c|c|c}
        \hline
        \multirow{1}{*}{Model} & \multirow{1}{*}{Gigaspeech-dev} & \multirow{1}{*}{Gigaspeech-test} \\
        \hline
        RNN-T & 14.7 & 14.5 \\
        FNT & 14.8 & 14.6 \\
        C-FNT & 14.7 & 14.5 \\
        \hline
    \end{tabular}
    \vspace{-0.4cm}
\end{table}

\subsection{Results}
Before assessing the effectiveness of our proposed model, we first examine whether the incorporation of class-based LM in FNT affects its original performance. Therefore, an empty name list is used for C-FNT. The WER results on the standard dev and test sets of gigaspeech are reported in table\ref{empty name list test}. It can be seen that C-FNT, RNN-T and FNT yield similar WER performance on the general ASR task. This suggests that C-FNT is able to retain its original performance, allowing us to focus on evaluating its effectiveness when given a name list.

With the use of a name list and the same beam size, as indicated by the 3rd row in Table \ref{tab:wer_ner_c_fnt}, C-FNT boosts the NER performance by a large margin, yielding 7.2\% and 7.6\% relative gain in overall WER, 27.9\% and 30.8\% relative gain in terms of F1-score on Giga-name-test and NER-name-test respectively. This indicates that our proposed C-FNT has really learned the name class, and the tailored decoding algorithm can effectively decodes name from the class. At last, we use dynamic beam size and expect that promising beam would not be pruned during decoding in the name class. As indicated in the last row of Table \ref{tab:wer_ner_c_fnt}, the results match our expectation with a slight increase in terms of ASR and NER performances.

\vspace{-0.1cm}
\section{Conclusion}
\vspace{-0.1cm}
In this paper, we aim to improve the named entity recognition in E2E models. Unlike most previous work based on contextual biasing approaches, we proposed a novel model named C-FNT by integrating the class-based LM and factorized neural Transducer. The beam search decoding algorithm is tailored for C-FNT to facilitate named entity recognition. According to our experimental result, significant error reduction in named entities is achieved without hurting the performance of common words. Another prominent advantage of our proposed C-FNT is its flexibility and modularization. The class-based LM can be updated or extended to support other types of named entities or domains.

\section{Acknowledgements}
\vspace{-0.15cm}
This work was supported by the National Natural Science Foundation of China  (No. 62206171 and No. U23B2018), Shanghai Municipal Science and Technology Major Project under Grant 2021SHZDZX0102, the International Cooperation Project of PCL and Alibaba Innovative Research Program.

\bibliographystyle{IEEEtran}
\bibliography{mybib}

\end{document}